\PassOptionsToPackage{table,svgnames}{xcolor}
\documentclass[sigconf]{acmart}
\usepackage{graphicx} 
\usepackage{subcaption}
\usepackage{amsthm,amsmath,amssymb}
\usepackage{amsfonts}
\usepackage{color}
\usepackage{algorithmic}
\usepackage{bbm}
\usepackage{mathtools}
\usepackage{comment}
\usepackage{longtable}

\usepackage[ruled, linesnumbered]{algorithm2e}
\SetArgSty{textnormal}

\usepackage{multirow}

\newtheorem{assumption}{Assumption}
\usepackage{soul}
\usepackage[normalem]{ulem}

\AtBeginDocument{%
  }

\setcopyright{none}
\renewcommand\footnotetextcopyrightpermission[1]{} 
\def\phcomments{1}

\newcounter{note}[section]
\renewcommand{\thenote}{\thesection.\arabic{note}}
\newcommand{\note}[2]{\refstepcounter{note}\marginpar{\small\bf \textcolor{red}{#1~\thenote}}$\ll${\sf \textcolor{red}{#1's
 Comment~\thenote:}} {\em \textcolor{red}{#2}}$\gg$}

\ifnum\phcomments=1
\newcommand{\notePH}[1]{\note{PH}{#1}}
\newcommand{\notePM}[1]{\note{PM}{#1}}
\newcommand{\noteCS}[1]{\note{CS}{#1}}

\newcommand{\noteCN}[1]{\note{CN}{#1}}
\else
\newcommand{\notePH}[1]{}
\newcommand{\notePM}[1]{}
\newcommand{\noteCN}[1]{}
\newcommand{\noteCS}[1]{}

\fi

\begin{document}

\title{Optimal Workload Placement on Multi-Instance GPUs}

\author{Bekir Turkkan}
\affiliation{%
  \institution{IBM Research}
  \country{United States}
}
\author{Pavankumar Murali}
\affiliation{%
  \institution{IBM Research}
  \country{United States}
}
\author{Pavithra Harsha}
\affiliation{%
  \institution{IBM Research}
  \country{United States}
}
\author{Rohan Arora}
\affiliation{%
  \institution{IBM Research}
  \country{United States}
}
\author{Gerard Vanloo}
\affiliation{%
  \institution{IBM Research}
  \country{United States}
}
\author{Chandra Narayanaswami}
\affiliation{%
  \institution{IBM Research}
  \country{United States}
}

\renewcommand{\shortauthors}{Turkkan et al.}

\begin{abstract}
    There is an urgent and pressing need to optimize usage of Graphical Processing Units (GPUs), which have arguably become one of the most expensive and sought after IT resources. To help with this goal, several of the current generation of GPUs support a partitioning feature, called Multi-Instance GPU (MIG) to allow multiple workloads to share a GPU, albeit with some constraints. In this paper we investigate how to optimize the placement of Large Language Model (LLM)-based AI Inferencing workloads on GPUs. We first identify and present several use cases that are encountered in practice that require workloads to be efficiently placed or migrated to other GPUs to make room for incoming workloads. The overarching goal is to use as few GPUs as possible and to further minimize memory and compute wastage on GPUs that are utilized. 
    We have developed two approaches to address this problem: an optimization method and a heuristic method. We benchmark these with two workload scheduling heuristics for multiple use cases. Our results show up to 2.85x improvement in the number of GPUs used and up to 70\% reduction in GPU wastage over baseline heuristics. We plan to enable the SRE community to leverage our proposed method in production environments.
\end{abstract}

\maketitle
\pagestyle{empty}

\section{Introduction}
\label{s:introduction}

Generative AI powered by large language models (LLMs) marks the beginning of a new era in improving human productivity.
While current applications range from email assistants and conversational interfaces to systems that assist software engineers with code generation, significantly more sophisticated applications are rapidly anticipated.
More recently, LLM-based multi-agent systems have achieved considerable progress in complex problem-solving~\cite{guo2024largelanguagemodelbased}.
GPUs are at the heart of LLM pretraining, fine-tuning, and inferencing.
They have become one of the most prized and expensive resources in computing with demand far exceeding availability.
Energy consumed by AI and GPUs in particular have become a cause for concern as well. For example, the International Energy Agency (IEA) estimated that data centers, cryptocurrencies, and artificial intelligence (AI) consumed about 460 TWh of electricity worldwide in 2022.
This number is expected to grow to 1000 TWh in 2026 \cite{IEAElectricityReport}.
GPU optimization has thus become critical for efficient use of limited resources and energy.

The AI community is working on LLM optimizations at various levels, including structuring and pipelining training computations, leveraging speculative execution for inferencing using a mixture of models,  building smaller purpose-specific models and multi-modal models, leveraging quantization to create lower precision models, optimizing queue management for LLM serving, and so on. Usage patterns and applications for LLMs are also rapidly evolving, spanning even synthetic data generation and automated evaluation of LLM inferencing.  Every LLM developer\textemdash including OpenAI, Meta, Microsoft, IBM, Anthropic, and Mistral\textemdash has released models whose sizes currently range from roughly 3 billion parameters to 70+ billion parameters.

Simultaneously, GPU designers realized that these smaller models can fit into a partitioned GPU and have begun providing access to fractional partitions of GPUs instead of just whole units.
This enables the hosting of multiple models or multiple instances of the same model on a single GPU.
We expect this trend to continue.
While Nvidia is the leading provider of GPUs today, competitors like AMD, Intel, and others will apply similar approaches to GPU partitioning.

\begin{figure}[h]
  \centering
  \includegraphics[width=0.9\linewidth]{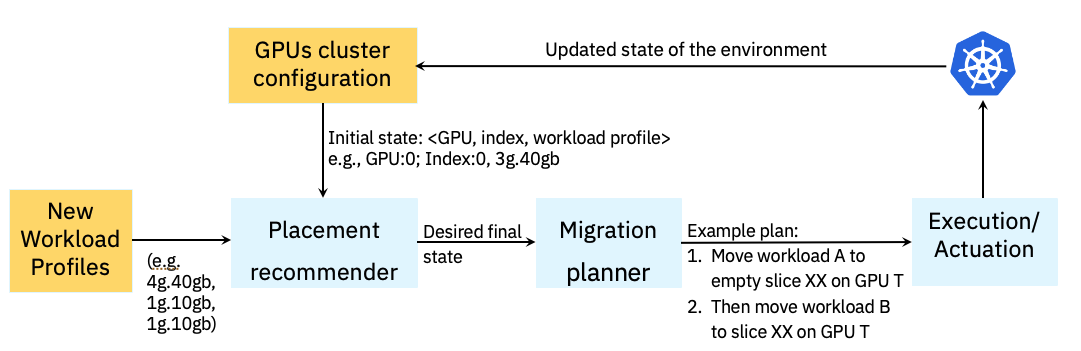}
  \caption{Framework Overview}
  \label{fig:framework}
\end{figure}

Smaller models coupled with the ability to use fractional partitions on GPUs have created an opportunity to better utilize the GPUs available and maximize the availability of resources for future workloads.
With inferencing making up the vast majority of these LLM-based AI workloads, we propose a framework depicted in Figure ~\ref{fig:framework} to achieve the optimal placement of such inferencing workloads. This framework can be extended to support fine-tuning of LLMs as well, and consists of the following main components:
\begin{enumerate}
    \item \textit{Placement Recommender:} This component determines the GPU and the partition on which new and existing LLM inferencing workloads should be placed. It identifies a desired final state from the current GPU cluster configuration. 
    \item \textit{Migration planner}: This component determines a migration plan that includes the movement of existing workload from their initial state, placement of new workload and any sequential aspects that are associated with these moves to produce the final state.
    \item \textit{Executor/actuator}: Implements the migrations to achieve the desired final state.
\end{enumerate}

This paper focuses on the design of intelligent placement recommenders.  In Section 2, we identify multiple motivating use cases (initial deployment, compaction and reconfiguration) with associated examples and performance metrics to highlight the benefit of efficient placement of models on MIGs  on several key performance metrics. In Section 3, we formally lay out the multiple objectives and constraints towards the design of such an intelligent recommender. We note that one of the goals of our recommender, besides maximizing GPU utilization, is to minimize sequential migration, i.e., dependence on other workloads to achieve desired state. 

The main contribution of our work is in proposing two approaches to solve the different use cases where both consider the same constraints and objectives (see Section 4). The first approach is a novel multi-objective global optimization approach using mixed-integer programming (MIP) that simultaneously handles all the use cases. The second approach is a fast rule-based heuristic that addresses the use cases separately. At the core, the approaches inherently have a bin-packing flavor to them and because this is a NP hard problem, the computational overhead of the exact method can become significant for large sized clusters. The main differences from a traditional bin-packing problem to the placement problem are that there is a notion of an index for a workload in a bin/GPU and that workloads can occupy only contiguous GPU slices. Moreover, certain workloads in certain indices can result in unusable parts of the GPU (wastage). The preferred method of choice besides performance on the KPI metrics are the latency requirements and the amount of sequential migration overhead that is needed. In Section 5, we  compare our proposed approaches with two other baselines approaches with comprehensive evaluations, which also include the need for sequential migration. The results show that the MIP method when used to jointly perform initial placement, compaction and reconfiguration, shows up to 6 - 11\% improvement in the number GPUs used, and reduces compute wastage  (up to 41\%) and increases availability (up to 10\%) than the other approaches. In compaction, MIP has up to 11\% improvement over heuristic-based approaches in terms of GPUs used, and up to 40\% reduction in wastage. Both of our approaches also outperform (39 - 65\% improvement in number of GPUs, and 40\% reduction in wastage) over other heuristics in the reconfiguration case. Our rule-based heuristic approach has competitive performance with MIP approach and requires less computational overhead for large-scale optimizations.

\section{Background and Motivation}
\label{s:background}
The Multi-Instance GPU (MIG) feature introduced by Nvidia allows GPUs to be partitioned into smaller fractions (currently a maximum of seven separate GPU instances).
This feature is supported by Nvidia's Ampere, Hopper, and\textemdash more recently\textemdash Blackwell GPUs~\cite{nvidia-mig}.
In this work, we focus on optimal placement of AI model deployments, especially small LLM deployments for inference tasks.
These are commonly deployed in containerized environments deployed and managed with Kubernetes.   
In this section, we first define MIG terminology used throughout the paper. Then, we explain the current state of MIG deployment technologies.
Finally, we conclude with identifying the practical use cases that motivate MIG placement optimization.

\subsection{Definitions}
\label{s:definitions}

\textbf{Workload:} A deployment associated with an LLM-inferencing service\textemdash a service serving a LLM model that can be hosted within a MIG instance.\cite{miao2023efficientgenerativelargelanguage}
We consider every replica of a Kubernetes deployment or new models to be equivalent in terms of placement optimization and define each as a workload.

\begin{figure}[h]
  \centering
  \includegraphics[scale = 0.33]{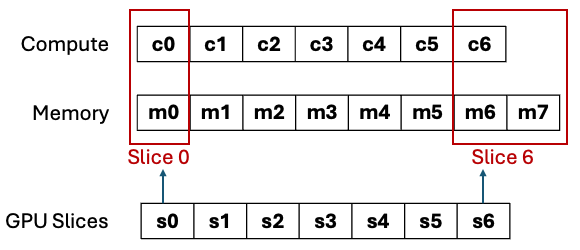}
  \caption{GPU Slices}
  \label{fig:slicing}
  \Description{}
\end{figure}

\textbf{Slice:} MIG offers dividing compute and memory resources of a GPU into fractions, called as a \textit{slice}.  Figure~\ref{fig:slicing} shows slicing for A100 and later generation GPUs.
These GPUs have 7 GPU slices, which corresponds to 7 compute and 8 memory slices. 
GPU slice 0 (s0) represents the first slice and includes the first compute (c0) and memory(m0) slices.
The additional memory slice (m7) can be used only with the last compute slice as part of GPU slice 6.

\textbf{Partition and Profile:} A partition is defined as a a partial GPU that includes one or more GPU slices. However, 
partitions are not fluid, and 
MIG requires using predefined set of partition profiles for each GPU model.
Nvidia supports 4 profiles for A30 and 7 profiles for A100 and later generation GPUs.
Each profile has a fixed number of compute and memory slices.
In this work, for illustrations and our experiments, we use GPUs with 7 profiles.
A100 and later generations use the same number of compute and memory slices for each profile.
Their corresponding memory size and compute power differ depending on the total GPU memory and the number of streaming multiprocessor (SM) blocks. For example, Nvidia H100-96GB GPU model has 12 GB memory for the smallest profile, while A100-80GB model has 10GB.
A partition is denoted by the compute slices and memory it is allocated: 2g.20gb refers to using 2 compute slices on the GPU and 2 memory slices, which has 20gb of memory in total.
To deploy a workload on a MIG-enabled GPU, first a new partition needs to be created for the profile of the workload.
Then, workload can be deployed to the partition. Table~\ref{tab:profiles} shows the common profiles, the corresponding memory and compute slice counts, the number of GPU slices required for partitioning, and profiles for the Nvidia A100-80GB GPU.
It also lists the possible starting GPU slice index as allowed indexes, the ordering of which will be further explained in Section~\ref{s:indexes}.

\begin{table}
\scalebox{0.9}{
   \small
    \centering
    \rowcolors{3}{}{gray!10}
    \begin{tabular}{*6c}
        \toprule
         Profile & \multirow{1}{*}{Instance} & Compute & Memory & \multirow{1}{*}{\# of GPU}  & \multirow{1}{*}{Allowed}  \\
         Id & Profile & Slices & Slices & Slices & Indexes \\
         \midrule
          0 & 7g.80gb & 7 & 8 & 7 & [0] \\
         5 & 4g.40gb & 4 & 4 & 4 & [0] \\
         9 & 3g.40gb & 3 & 4 & 4 & [4,0] \\
         14 & 2g.20gb & 2 & 2 & 2 & [4,0,2] \\
         15 & 1g.20gb & 1 & 2 & 2 & [6,4,0,2] \\
         19 & 1g.10gb & 1 & 1 & 1 & [6,4,5,0,1,2,3] \\
         20 & 1g.10gb+me & 1 & 1 & 1 & [6,4,5,0,1,2,3] \\
         \bottomrule
    \end{tabular}}
     \caption{A100 MIG Instance Partition Profiles}
     \label{tab:profiles}
\end{table}

\textbf{Configuration:} Configuration can be defined as the information about existing partitions and their indexes on a GPU, along with workload assignments to the partitions. Thus, it shows the current state of a GPU in terms of partitions, their allocations, and availability. 

\textbf{Migration:} Due to optimization decisions, workloads can commonly be relocated to different GPUs or different partitions on the same GPU.
We define any workload-partition assignment change as migration. 
Migrations may need a workload to be drained from the current GPU partition to which it is allocated.
There is also a model loading and initialization time to be considered when the workload needs to be redeployed on a different GPU partition.
If a workload has to be drained from the current GPU partition before it becomes actively running on the target GPU partition, it causes downtime which we call as disruptive-migration. 
On the other hand, non-disruptive migration can be defined as the ability to create a replica of the model to migrate in the new partition without dropping user requests for this or any other model.
If destination partition is unoccupied, it can be achieved in one step.
But if it is occupied, then the migration has to be achieved sequentially in multiple steps.
To avoid any downtime and achieve non-disruptive migration, we assume that a replica of the workload may be created and placed optimally prior to the original workload being drained from the old GPU partition.
This assumption allows us to assume that the time for which a migrated model is ``unavailable'' is negligible and can be ignored.

\subsection{Deployments with Multi-Instance GPUs}
\label{s:mig}

As a rule of thumb, the memory required for deploying a model is at least 2 times the number of model parameters (3B parameter model can fit in a MIG instance with 10gb, 7B parameter model in 20gb, 34B parameter model in 80gb, etc)\cite{anyscale-numbers-every-llm-developer}.
The remaining memory is used as KV-cache to store generated tokens and mainly decides the number of concurrent inference requests to be handled.
The required cache size per token changes depends on the number of layers, dimension, and precision of transformer model and can roughly be calculated with Equation~\ref{eq:memory_token} \cite{nvidia-llm-optimization}.
In this regard, partitions with same amount of memory (such as 1g.20gb and 2g.20gb) can be considered to serve similar number of concurrent requests.
\begin{align}
\label{eq:memory_token}
cache\_size = 2 * num\_layers * dimension * precision
\end{align}
Section~\ref{s:gpuOperator} and ~\ref{s:DRA} summarizes the current state and developing features of common MIG technologies, respectively.

\subsubsection{Nvidia GPU Operation}
\label{s:gpuOperator}
MIG deployments commonly utilize the Nvidia GPU operator to deploy LLMs on Kubernetes because of its flexibility and scalability.
The operator partitions GPUs within a cluster based on a user-defined policy for each node.
As a consequence, this leads to all GPUs on a node being configured in the exact same way. Any changes to this policy requires restarting the already running workloads, which is very disruptive. 
For example, trying to deploy a new workload with a 2g.20gb profile on a node with a Hopper 100 80 GB GPUs partitioned into 4g.40gb and 3g.40gb slices will result in the new deployment being stuck in a ``pending'' state since there is no available partition of the same size. 
In order to meet the demands of serving many different models, clusters will need to have diverse partitioning strategies.
However, these limitations can be resolved with dynamic resource allocation, which we describe next.
\subsubsection{Dynamic Resource Allocation}
\label{s:DRA}
Dynamic Resource Allocation (DRA) is a Kubernetes feature\textemdash currently under development\textemdash that allows flexible resource scheduling~\cite{DRA}.
DRA aims to provide full flexibility in creating and deleting partitions, converting new GPUs to MIG mode, and allowing differently configured MIG partitions for each GPU on the same node. Importantly, it supports these changes in a non-disruptive way, such that the currently running workloads are not affected by the changes in other slices of the same GPU or others on the same node.

The work presented in this paper therefore focuses on developing a workload placement framework to optimize GPU utilization and cost efficiency using DRA.

\subsection{Placement Optimization Use Cases}
\label{s:usecases}
We consider three commonly encountered use cases for placement optimization of MIG instances: \textit{initial deployment}, \textit{compaction}, and \textit{reconfiguration}. We describe each of them below.

\subsubsection{Initial Deployment}
\label{s:initialDeployment}
\begin{figure*}
\centering
\begin{subfigure}[t]{0.45\linewidth}
  \centering
  \includegraphics[width=\linewidth]{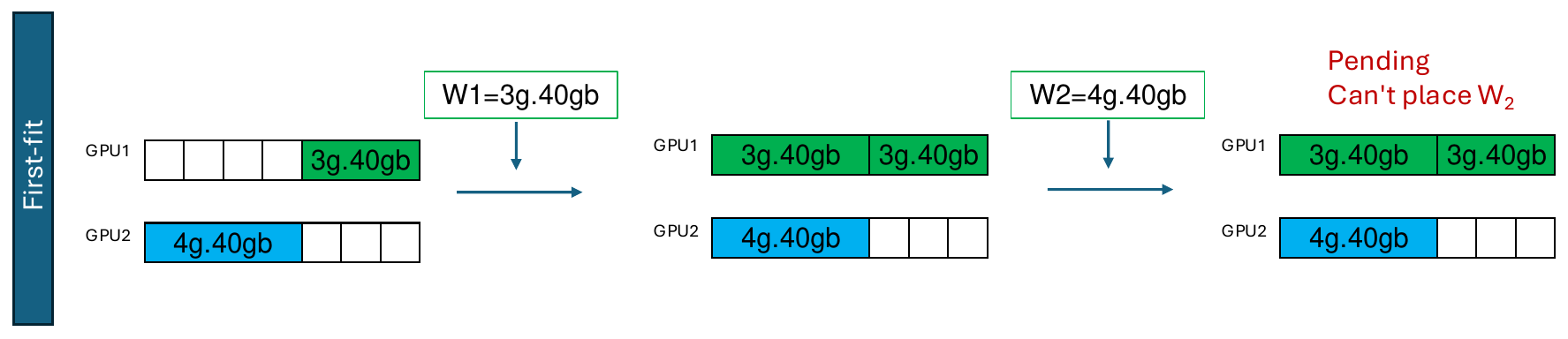}
  \caption{First-fit placement}
  \label{fig:first-fit}
\end{subfigure}
\hfill
\begin{subfigure}[t]{0.45\linewidth}
  \centering
  \includegraphics[width=\linewidth]{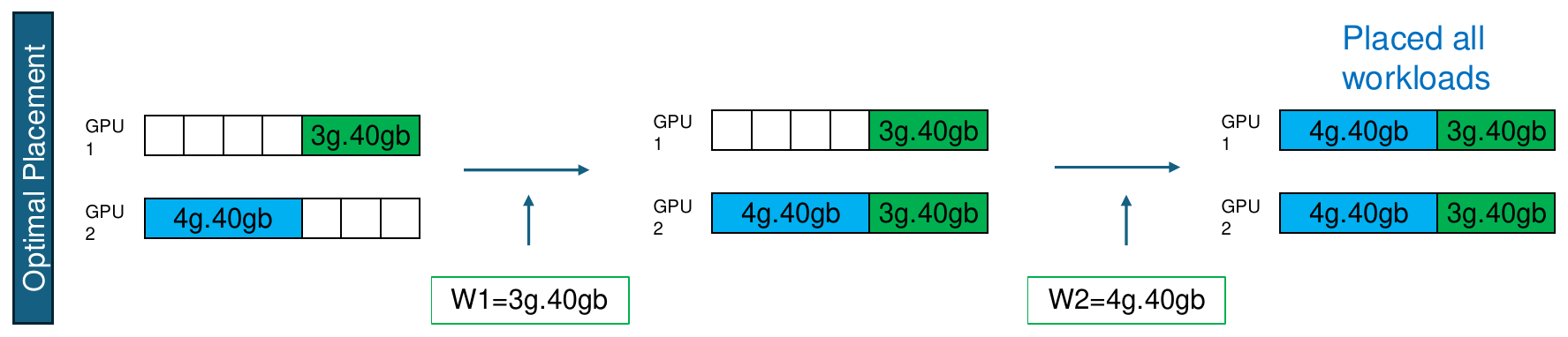}
  \caption{Optimal placement}
  \label{fig:optimal-placement}
\end{subfigure}
\caption{Initial deployment of new workloads}
\label{fig:initial}
\end{figure*}

Based on the incoming request load, LLM-based inferencing workloads are scaled to add or remove replicas.
\textit{Initial deployment} refers to partitioning a GPU to use one or more GPU slices and then placing a new workload in it. Due to the dynamic nature of task diversity and incoming request load, deploying new workloads on the GPU is a frequent event. Thus, efficient placement of these models will ensure better utilization, fewer disruptions due to placement changes, and greater availability of GPUs for future incoming deployments.
Figure~\ref{fig:initial} shows a simple first-fit placement based on device IDs and an optimal placement with regards to minimizing resource wastage.
It starts with an initial state that has 2 GPUs and 1 existing workload running on each GPU. Then, a new workload $w_1$ with 3g.40gb profile needs to be placed. The first-fit algorithm (top of Figure~\ref{fig:initial}) finds the first GPU available and puts $w_1$ on GPU1, which uses 4 GPU slices (4 compute and 4 memory slices) and wastes a compute slice.
Once a second workload $w_2$ with 4g.40gb profile needs to be placed, the algorithm is unable to find a partition to fit the new workload.
Thus the workload is stuck in the \textit{pending} state. However, an optimal algorithm (bottom of Figure~\ref{fig:initial}) that considers GPU resource wastage would place the $w_1$ on GPU2 and uses 3 GPU slices (3 compute and 4 memory slices) to avoid compute slice wastage. Then, it can successfully place the $w_2$ on GPU1 since it has availability for a 4g.40gb profile.

\subsubsection{Compaction}
\label{s:compaction}
\begin{figure}[h]
\centering
\includegraphics[scale=0.3]{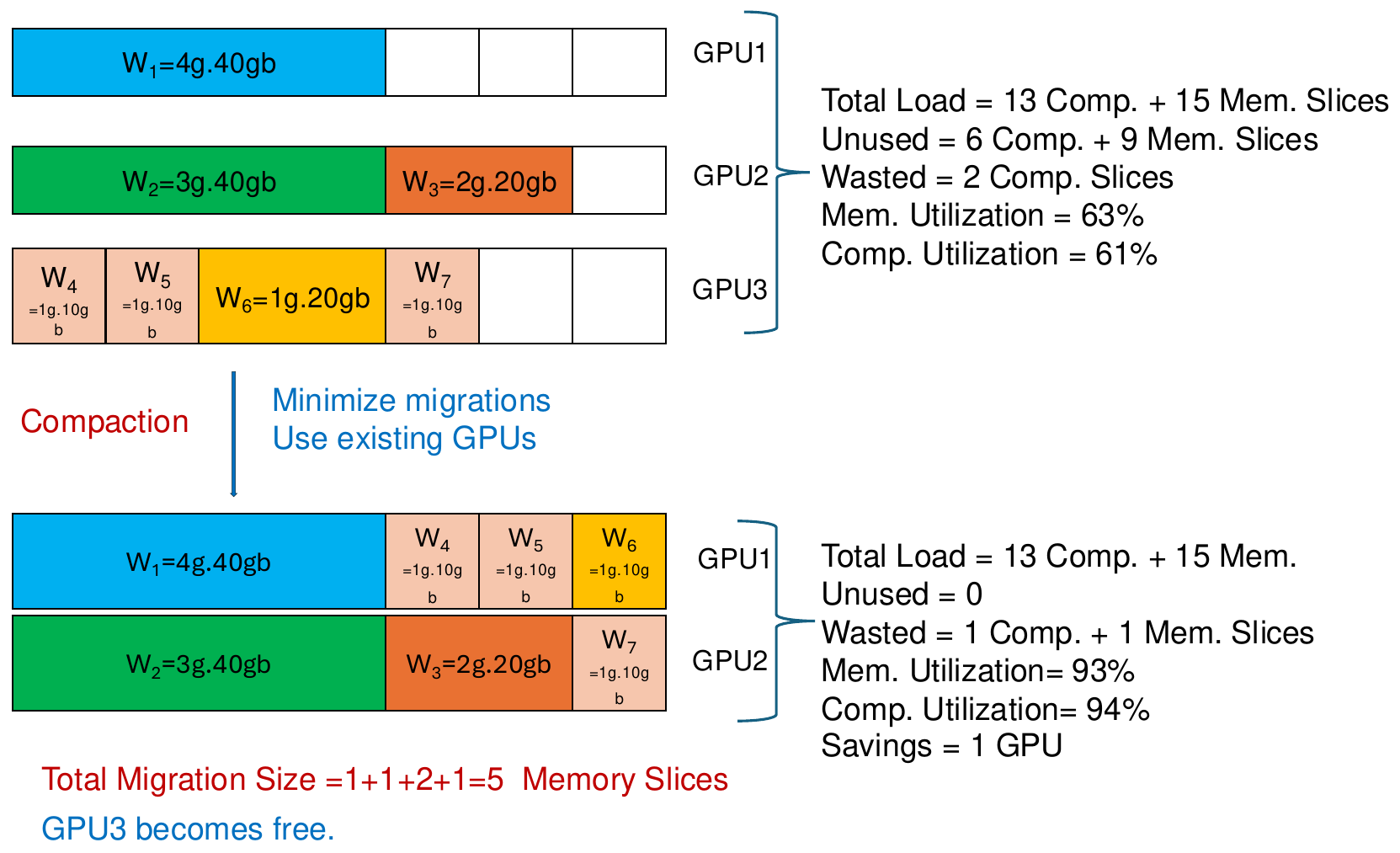}
\caption{Initial deployment or compaction}
\label{fig:compaction}
\end{figure}

Models are scaled in response to varying incoming request loads. When they complete their tasks and are no longer in use, they should ideally be removed to make room. Alternatively, if there is more demand than allocated resources, more replicas must be placed onto the GPU. These frequent adjustments can easily cause GPUs to become fragmented and underutilized. Therefore, efficient placement of existing models requires periodic adjustments to avoid under-utilizing the GPUs and allow for high availability of GPU resources for future workloads. We define compaction as vacating underutilized GPUs by migrating existing workloads to other GPUs in the cluster. The frequency of compaction can be adjusted based on how dynamic the changes in the workloads are. 
Figure~\ref{fig:compaction} shows an initial state of a node with 3 GPUs with some workloads. Initially, the compute and memory utilization of the 3 allocated GPUs is 61\% (13 slices out of 21) and 63\% (15 slices out of 24), respectively, due to 6 unused compute, 9 unused memory slices, and 2 wasted compute slices ($w_2$ and $w_6$ being the culprits). GPU1 and GPU2 have availability for 3g.40gb and 1g.20gb, respectively. Migrating workloads in GPU3 to GPU1 and GPU2 will free the resources in GPU3. Thus, compaction decreased the number of allocated GPUs by 1, and after compaction, the compute and memory utilization of allocated GPUs became 94\% (13 slices out of 14) and 93\% (15 slices out of 16) with no unused slice and only 1 compute slice ($w_2$) and 1 memory slice ($w_7$) wasted. 
This process requires moving workloads in size of 5 memory slices, which we denote as \textit{migration size}. 

\subsubsection{Reconfiguration/Redeployment}
\label{s:reconfiguration}
\begin{figure}[h]
\centering
\includegraphics[scale=0.3]{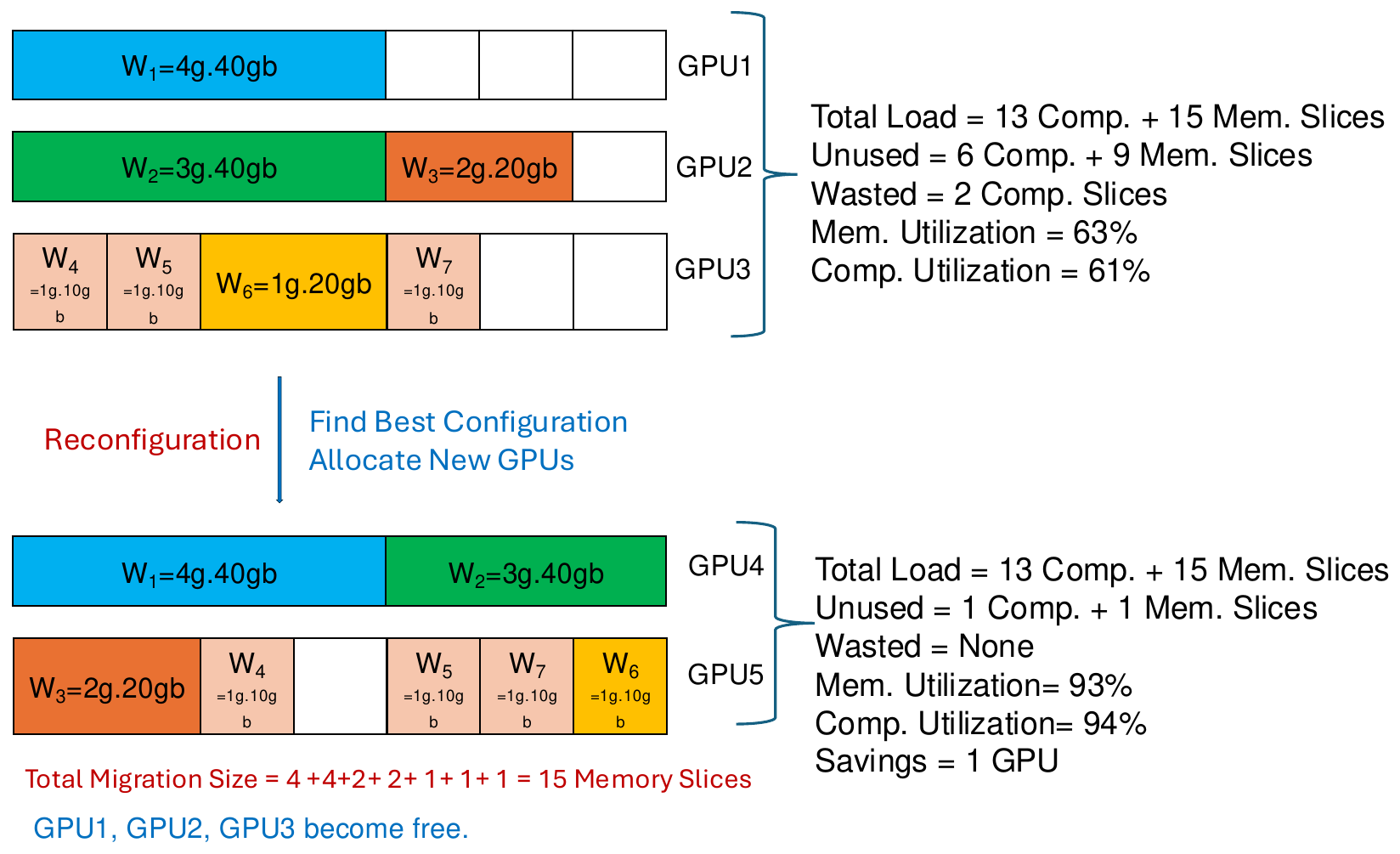}
\caption{Reconfiguration}
\label{fig:reconfig}
\Description{}
\end{figure}

Can we do better in the above scenario and avoid the wastage of 1 compute slice ($w_2$) and 1 memory slice ($w_7$)? Indeed, we can do so by reconfiguration or redeployment.  In this process we migrate all workloads from existing GPUs to new or unused GPUs, optimizing the placement of each workload in the new space. 
While it has a higher cost compared to the other cases due to the migration effort, it achieves ideal placement of all existing workloads in a non-disruptive way. New replicas are created on the GPUs and\textemdash once running\textemdash the existing ones are removed. Reconfiguration is preferred mainly in two cases: as part of maintenance cycles and node or provider change. During a maintenance or upgrade cycle, if a cluster has enough free GPUs to migrate all workloads, an optimal placement model can vacate underutilized GPUs and minimize wastage at the same time.
Figure~\ref{fig:reconfig} shows the same initial state as Figure~\ref{fig:compaction}. However, instead of migrating only the workloads on GPU3, it migrates all workloads to new GPUs. With reconfiguration, the GPU requirement is reduced by 1 GPU, while the wastage of compute or memory slices is completely avoided. In addition to placing all workloads in 2 GPUs, reconfiguration also provides an additional availability for another workload with 1 slice requirement. 

\textbf{Discussion:} Compaction and reconfiguration both target reducing the number of GPUs. Compaction uses already allocated GPUs, while reconfiguration leverages free GPUs to achieve non-disruptive migration. 

For example, consider achieving optimal state in Figure~\ref{fig:reconfig} using GPU1 and GPU2. It requires multiple steps to complete migration, as $w_2$ must be migrated to GPU1. Then, $w3$ must be relocated at GPU2 at index 0. Finally, $w_4, w_5, w_6$, and $w_7$ must be migrated to GPU2. Having free GPUs lifts the need for sequential migration steps and decreases the required time to achieve target state since it allows all migrations to be run simultaneously in one step. 
Besides sequential migration (which one may argue is non-disruptive), free GPUs also allow the system to achieve an optimal desired state in a non-disruptive fashion (i.e., without making the model unavailable).
Consider, for example, the case when we want to move from the target state in Figure~\ref{fig:compaction} and achieve optimal state in Figure~\ref{fig:reconfig} with only GPU1 and GPU2 at hand and no other. This is not possible without a disruption.

\section{Efficient Placement with MIG under DRA}
\label{s:efficientGPU}

\subsection{Objectives of MIG Placement}

\subsubsection{Objective 1: Minimizing number of GPUs.}
\label{s:minimizeGPU}
GPU compute requirements for workloads or models are varied.  For instance, hosting an LLM with more than 70B parameters require multiple GPUs while a model with 7B parameter can easily be deployed on a small MIG instance. 
Companies either have their private GPU clusters where they have limited GPU resources or they use public cloud and pay per their usage. For the private cluster case, minimizing the number of GPUs used 
provides flexibility for future workloads in two ways: (1) ability to use a GPU in a single GPU mode for larger workloads and (2) enabling MIG when needed for all possible partition profiles.
For the public cloud case, minimizing number of GPUs decreases the cost automatically. 

\subsubsection{Objective 2: Minimizing GPU compute or memory wastage.}
\label{s:minimizeWaste}
Current MIG slicing allows seven compute and eight memory slices for Ampere (A100), Hopper (H100, H200), and Blackwell GPU models.
MIG supports vertical slicing where each compute slice uses a memory slice. However, there are there profiles that use an additional memory slice, profile 0 (e.g., 7g.80gb for H100-80GB model), profile 9 (e.g., 3g.40gb for H100-80GB model) and profile 15 (e.g., 1g.20gb for H100-80GB model). 
For a MIG enabled GPU, if the last slice is not partitioned for these profiles, a memory slice is wasted. Similarly, profiles with uneven GPU and memory slices, 3g.40gb or 1g.20gb, may lead to wasting a GPU slice if they are not placed to include the last slice. 
An optimal placement model shall consider which slices to be used for partitioning of a given profile to minimize the waste or maximizing the allocation of slices.

\begin{figure}[h]
\centering
\includegraphics[width=\linewidth]{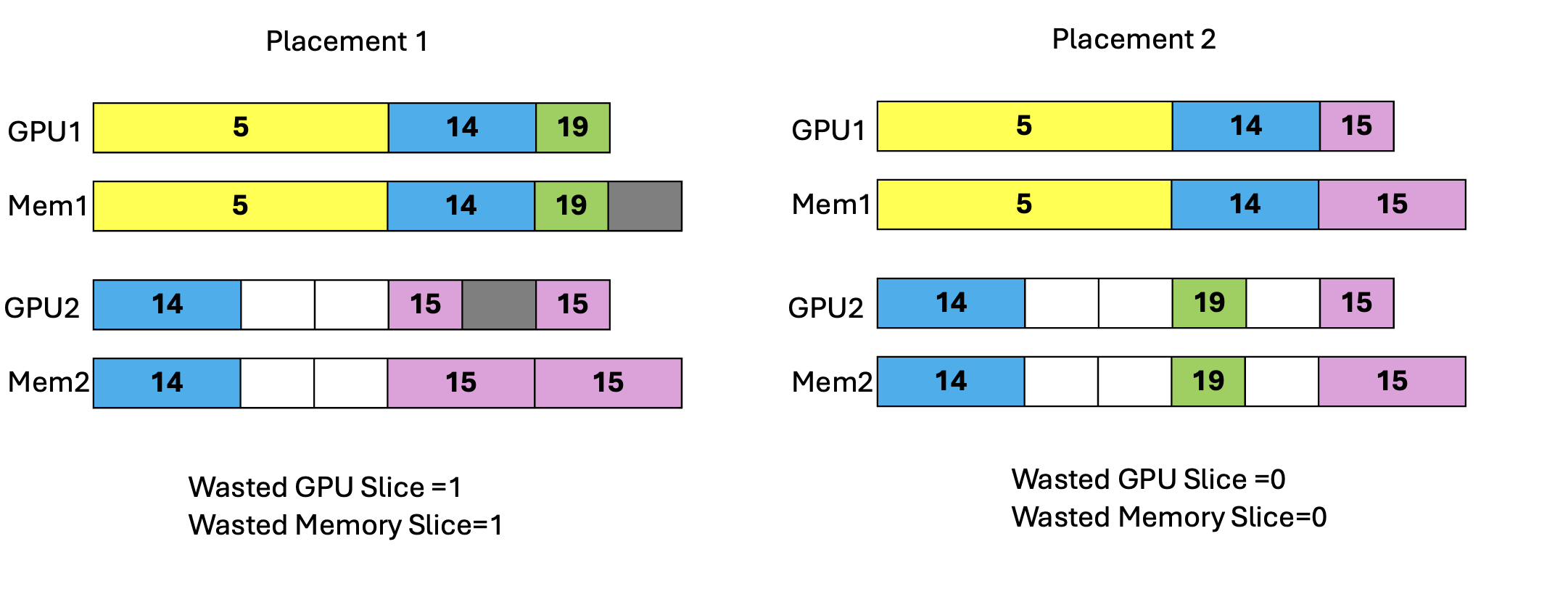}
\caption{Placement examples to show possible GPU waste.}
\label{fig:gpu-waste}
\Description{}
\end{figure}

Figure~\ref{fig:gpu-waste} shows 2 placement decisions for a given set of workloads with profiles as follows: 4g.40gb (Profile 5), 2gb.20gb (Profile 14), 2g.20gb, 1g.10gb (Profile 19), 1g.20gb (Profile 15), 1g.20gb. Placement 1 and placement 2 both use the same number of GPUs. However, placement 1 leads to wasting 1 GPU slice and 1 memory slice. Similarly, it results in 2 available slices for future workloads. On the other hand, placement 2 leads to no wastage and results in 3 available slices for future placements. Placement 2 targets minimizing GPU or memory waste and improves utilization of GPU/memory slices. To achieve this, we can either target minimizing GPU/Memory waste or maximize utilization (the total number of used memory and GPU slices).

\subsubsection{Objective 3: Maximizing GPU availability.}
\label{s:maximizeAvailability}
MIG instances can only be placed certain indexes based on their profiles as explained in Section~\ref{s:indexes}.
Availability of a GPU for future workloads highly depends on the underlying placement algorithm. Figure below shows a placement on A100-80GB gpu for a workload ($w_1$) with 3g.40gb requirement. This profile can be placed only at indexes 0 and 4. When it is placed at index 0, it uses the first 4 slices and for the remaining 3 slices can be used for another 3g.40gb, 2g.20gb, 1g.20gb or 1g.10gb. However, scheduling a 4g.40gb workload to this GPU after the initial placement would cause pending state and schedule would wait until a proper slice becomes available. On the other hand, placing the workload at index 4 prevents this issue and allows many different combinations to be placed for the remaining of the GPU. Therefore, a placement model shall target maximizing the number of consecutive available slices during placement of a single workload to increase the availability of GPU resources for future workloads.

\subsection{Constraints of MIG Placement}

\subsubsection{Constraint 1: MIG uses vertical slicing.}
\label{s:verticalSlicing}

A GPU partition can be created by using different number of GPU slices. 
Figure~\ref{fig:slicing} shows how vertical slicing works and dependency between compute and memory slices. 

\subsubsection{Constraint 2: Partition profiles can be created only on certain indices.}
\label{s:indexes}
MIG instances requires partitioning GPUs using a fixed set of profiles. Table~\ref{tab:profiles} shows the list of profiles for A100 and later generation Nvidia GPUs and their corresponding number of compute and memory slices. Nvidia drivers allow partitions to be created only at the listed indices. 
Table~\ref{tab:profiles} also shows the preference order for the indexes to create a partition with the corresponding profile on a MIG enabled GPU. For instance, a partition with profile 9 (3g.40gb) can be created either at slice 0 or slice 4. Preference order shows that creating it at slice 4 better than slice 0 for efficiency since when created at slice 0, it causes a compute slice to be wasted. To find the preference order for efficiency, we created every possible partition configuration through Nvidia-smi and record the slice indices for each partition. We empirically found the preference order in Table~\ref{tab:profiles} which also targets maximizing efficiency.

\subsubsection{Constraint 3: Additional memory slice can only be used with last slice of a MIG enabled GPU}
\label{s:additionalMemory}
MIG enabled GPUs have one additional memory slice that can be used with a partition including the last slice of the GPU as shown in Figure. In this regard, profile 0, 9, and 15 enable using additional memory slice when they include the last slice in their partition. 
Other profiles are not able to utilize the additional memory slice and cause memory waste.

\subsubsection{Constraint 4: Changing allocated slices of a GPU instance requires repartitioning.} 
\label{s:repartitioning} 
MIG drivers do not allow changing the number of compute or memory slices without deleting the partition to be modified and repartitioning it with the requested resources. 
Similarly, merging or splitting instances are not allowed without repartitioning.
Although time required for partitioning a GPU is small, it may lead to interruptions if there are any running workloads on the partitioned slices.

\section{Proposed Approaches for Efficient Placement}
\label{s:model}

In this section, we present two approaches to solving the placement and migration problem under DRA that spans the three use cases described in Sec.~\ref{s:usecases}. The first approach leverages mixed-integer programming to identify a global optimal solution by jointly optimizing across the objectives and constraints discussed in Sec.~\ref{s:efficientGPU}. 
The second approach is a rules-based heuristic that tackles each of the use-cases separately which is unlike the former method. Besides a decrease in computation overhead in the latter, the former solution may require a sequential migration policy to achieve the desired optimal state, while the latter is specifically designed so that migration is executed in one shot (i.e., there is no dependency of another workload migrating).

\begin{assumption}
 As workloads can be matched with a fixed number of discrete and finite profiles, we assume that every feasible combination of workloads for a GPU, i.e., one that satisfies bin packing constraints across its resource dimensions, can be permuted to match at least one feasible GPU partition profile.
\end{assumption}

This assumption allows us to view the optimal placement problem in general as a multi-dimensional bin packing problem across GPU resource dimensions without explicitly considering GPU placement indices (which are provided in Table~\ref{tab:profiles} for the A100 GPU), which can then be followed by an indexing step. We validate this assumption for the Nvidia-smi across A100 and H100 GPU types by exhaustively considering all permutations of workloads on a GPU that satisfy the bin-packing constraints along resource dimensions (i.e., memory limit of 80gb, compute limit of 7g and at most one {\em{me}}). However, we observe that in the presence of pre-existing workloads which are not movable, this assumption fails to hold because the partition profiles can be created only on specific indices, as discussed in constraint~\ref{s:indexes}. 
These partition-index constraints restrict the options available for ordering and partitioning of the un-partitioned slice of a MIG into smaller slices. 
We illustrate this with an example below. 

\begin{figure}[h]
\centering
\includegraphics[scale=0.35]{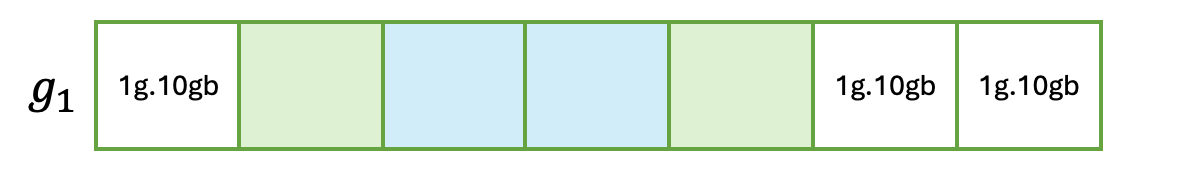}
\caption{GPU partitioning with DRA}
\label{fig:DRA1}
\Description{}
\end{figure}

Consider GPU $g1$ shown in Figure~\ref{fig:DRA1} where none of the placed workloads are movable. It has 3 1g.10gb workloads in indices 0, 5 and 6 with a 4g.40gb slice empty in between. Per the alignment feasibility constraints, only one of the following three combinations of workloads may be placed in the empty slice: (1) 1 workload of 2g.20gb at index 2 and 2 workloads of 1g.10gb at indices 1 and 4, (2) 1 workload of 1g.20gb at index 2 and 2 workloads of 1g.10gb at indices 1 and 4, (3) 4 workloads of 1g.10gb. It is infeasible to have a 4g.40gb or 3g.40gb in any free index, or 2g.20gb or 1g.20gb workloads at index 1 or 3. This means that the feasible slices that can be treated as a bin are: 1g.10gb slice at index 1 and 4, and a 2g.20gb slice at index 2. The 2g.20gb slice can accommodate either a 2g.20gb workload (option 1), 1g.20gb workload (option 2), or two 1g.10gb workloads (option 3). 

To address this above concern, we perform data preprocessing over all the GPUs with pre-existing workloads that we denote by set $\mathcal{G'}$ (which in other words is the set of all GPUs that are partially partitioned such as $g1$ in Figure~\ref{fig:DRA1}). The preprocessing detailed in Algorithm 1 outputs
a set $\mathcal{P}_g$ that represents the set of unallocated largest feasible partitions available on GPU $g \in \mathcal{G'}$ 
that can be re-partitioned. 
We define 
\[
\mathcal{P} = \mathcal{P}_1 \cup \mathcal{P}_2 \cup \mathcal{P}_3 \ldots \mathcal{P}_G \mbox{, where } G = |\mathcal{G'}|
\]
As per this definition, for $g1$ in Figure ~\ref{fig:DRA1}, $\mathcal{P}_{\mbox{\textbf{g1}}} = \{1g.10gb,2g.20gb,$ $ 1g.10gb\}$. Consider another GPU $g2$ where we have a 1g.20gb placed in last slice. Here pre-processing $\mathcal{P}$ would give $\mathcal{P}_{\mbox{\textbf{g2}}} = \{4g.40gb, $ $ 2g.20gb\}$. But in fact, we can just work with a ``merged'' set $\mathcal{P}_{\mbox{\textbf{g2}}} = \{6g.60gb\}$. A merged set is preferable since it reduces the number of variables in the optimization problem. The only time we do not work with a merged set if all of the large jobs cannot be placed in some index of the slots. Algorithm 1 does not incorporate this merging idea to keep the exposition of the pre-processing simple.

\begin{small}
\SetKwInOut{Parameter}{Parameter}
\begin{algorithm}[!htb]
    \caption{Determine $\mathcal{P}_g$}
    \KwIn{$\mathcal{I}$: Sorted list of profile IDs by GPU slice size; \hspace{1.5cm} $\mathcal{K}$:  Ordered list of GPU slice indices; \hspace{2.5cm} $g$: A partially partitioned GPU; }
    \Parameter{$c_i, m_i, s_i$ are the compute, memory and GPU slices required $\forall i \in \mathcal{I}$ }
          $\mathcal{P}_g \gets \emptyset$\;
          \For{$k \in \mathcal{K}$}{
                \If{index $k$ is not partitioned in $g$}{
                    \For{$i \in \mathcal{I}$}{
                       \If{type $i$ workload can be placed in slice $k$}{
                       Place this hypothetical load in $g$;\\
                       Add $\{c_i, m_i\}$ to $\mathcal{P}_g$ 
                       }
                    }
                }
          }
         \KwOut{Return $\mathcal{P}_g$}
\end{algorithm}
\end{small}

\subsection{Optimal Placement and Migration Using Mixed-Integer Programming (MIP)} 
\label{s:mipModel}
We model workload placement and migration on a heterogeneous GPU cluster based on the \textit{two-dimensional bin packing problem}, wherein each GPU 
is considered a bin that is empty (completely unpartitioned), or is partially partitioned with some workloads, or fully partitioned with existing workloads. Here we only consider the compute and memory dimensions. If there are other resource dimensions (e.g., like media extensions - {\em me}), the formulation below can be extended to a multi-dimensional bin-packing problem.

Let us consider a set of model instances or workloads $\mathcal{W}$ along with their optimal profiles (e.g., 1g.20GB), that is, compute instance requirement $c_i$ and memory requirement $m_i$ for each model $w \in \mathcal{W}$. We are also provided a set of unused GPUs $\mathcal{G}$ each of which can be partitioned in a pre-specified number of ways such that the total size across the partitions does not exceed the maximum allowed number of compute instances $C_g$ and the total memory does not exceed the GPU memory size $M_g$ for GPU $g \in \mathcal{G}$. We assume that the GPU operator used enables heterogeneous partitioning of GPUs in $\mathcal{G}$ and provides the ability to place a workload on a specific GPU $g \in \mathcal{G}$ as per the optimization solution.

Next, we present a mixed integer programming (MIP) model, \textbf{WPM}, that jointly handles initial placement of new workloads, migrating workloads that have already been placed on GPU slices and GPU reconfiguration. 
If free GPUs are not available, for modeling purposes, we consider adding imaginary GPUs, $\mathcal{G}^i$, such that every GPU $g' \in \mathcal{G'}$ has a corresponding imaginary GPU $g^i \in \mathcal{G}^i$, and only one of them appear in the final solution. In our model, if we use an imaginary GPU, it is equivalent to repartitioning the original GPU $g'$. For ease of notation, going forward we assume the set $\mathcal{G}^i$ is included in the set of unused GPUs $\mathcal{G}$.

The workload placement and migration problem (\textbf{WPM}) is formulated as a profit maximization problem and associates various actions with a revenue and a cost. These include the reward $p_w$ for placing workload $w$, the costs $q_g, \gamma^R_g, \gamma^W_g$ representing the cost of using GPU $g$, repartitioning GPU $g$ and wasting memory or compute slices in GPU $g$ respectively. Lastly,  cost $\gamma^M_w$ is the cost of migrating working across GPUs. 
\begin{assumption}
As the MIP is predominantly a bin-packing formulation that ignores placement indexing, the MIP approach inherently assumes that the migration cost within a GPU $g$ is 0. 
\end{assumption}
Although in practice this assumption may not be the case, migrating within a GPU is less expensive than across GPUs since model weights and cache can be transferred using the GPU memory. Assumption 2 also implies that the cost due to migrating a workload from its current GPU to its imaginary counterpart is also 0 as only one of the GPUs appears in the final solution. This is because using an imaginary GPU is representative of the GPU being repartitioned, a repartitioning cost is associated with such an action.

We present the \textbf{WPM} formulation next. This leverages the notation described in Table~\ref{tab:notation2}.

\begin{table}
\caption{Notation for \textbf{WPM}}
\label{tab:notation2}
\begin{center}
\scalebox{0.9}{
\begin{tabular}{rl}
\hline 
\multicolumn{2}{l}{\bf{Indices}}\\
$w$, & model instance\\
$g, g'$, & GPU index\\
\multicolumn{2}{l}{\bf{Sets}}\\
$\mathcal{W}$, & set of placed models that can be migrated and to-be\\&  placed models\\
$\mathcal{G}$, & set of free and imaginary GPUs.\\ 
$\mathcal{G'}$, & set of GPUs with preexisting workload\\ 
$\mathcal{G}^i$, & set of imaginary GPUs, $\mathcal{G}^i \subset \mathcal{G}$, introduced for every \\ & $g' \in \mathcal{G'}$ where all workloads on $g'$ can be migrated\\ 
$\mathcal{P}$, & set of free partitions in $\mathcal{G'}$,  $\mathcal{P} = \mathcal{P}_1 \cup \mathcal{P}_2 \cup \ \mathcal{P}_3 \ldots \mathcal{P}_{|\mathcal{G'}|}$\\
$\mathcal{A}$, & set of current model-GPU assignments, $(w,g') \in \mathcal{A}$ for \\
& placed models $w \in \mathcal{W}$ on $g' \in \mathcal{G'}$ \\
$\mathcal{B}$, & set of pairs of GPU and its imaginary counterpart $(g',g)$\\ &  where $g' \in \mathcal{G'}$, $g \in \mathcal{G}^i$.\\
\multicolumn{2}{l}{\bf{Parameters}}\\
$p_w$, & reward associated with placing workload $w$\\
$q_g$, & cost of using GPU $g$\\
$c_w$, & compute slice needed for model $w$\\
$m_w$, & memory needed for model $w$\\
$C_g$, & total compute slices on GPU $g$ ($C_g$ = 7 in A100)\\
$M_g$, & total memory size for GPU $g$ ($M_g$ = 80gb in A100, )\\ \vspace{0.05cm}
$S_g$, & common memory factor across GPU $g$ slices (10gb in A100)\\ \vspace{0.05cm}
$\gamma^M_w$, & penalty factor associated with migrating workload $w$\\ \vspace{0.05cm}
$\gamma^R_g$, & penalty factor associated with reconfiguring GPU $g$\\
$\gamma^W_g$, & penalty associated with wasted resources on GPU $g$\\

\multicolumn{2}{l}{\bf{Decision Variables}}\\
$x_{wg}$, & 
is 1 if model $w$ is assigned to GPU $g$, and 0 otherwise\\
$y_{g}$, & 
is 1 if GPU $g$ is used, and 0 otherwise\\
$z_g$, & auxiliary variable for partition $g \in \mathcal{P}$  that is $1$ when \\ &  a model is assigned to it, and $0$ otherwise\\
$u_g, v_g$, & slack in compute and memory resource constraints\\
$U_g, V_g$, & auxiliary variable corresponding to excess compute over\\ &  memory and when compute is full, the vice-versa\\
$\delta_g$, & auxiliary binary variable denoting if compute is full\\
\hline
\end{tabular}}
\end{center}
\end{table}

\begin{subequations}
\allowdisplaybreaks
\begin{align}
\begin{split}
\textbf{WPM}: \hspace{0.5cm}  & \\
\max_{\substack{x,y,z, \delta \in \{0,1\}\\ u,v, U,V \geq 0}} \;\; & \sum_{g \in \mathcal{G} \cup \mathcal{P}, w \in \mathcal{W}} p_w x_{wg} - \sum_{g \in \mathcal{G} \cup \mathcal{G'}} q_g y_g - \sum_{g \in \mathcal{G}^i} \gamma^R_g y_{g}  \\
& - \hspace{-.3cm} \sum_{\substack{(w,g') \in \mathcal{A},\\ (g',g) \in \mathcal{B}}} \gamma^M_w (1- x_{wg'} - x_{wg}) - \hspace{-.3cm} \sum_{g \in \mathcal{G} \cup \mathcal{P}} \gamma^W_g (U_g+ V_g) \label{eqn: 1a}
\end{split}\\[0.1cm]
\text{s.t. }\; & \sum_{w \in \mathcal{W}} x_{wg} \leq C_g y_g \quad \forall g \in \mathcal{G} \label{eqn: 1b}\\[0.1cm]
& \sum_{w \in \mathcal{W}}x_{wg'} \leq C_{g'}z_{g'} \quad \forall g' \in \mathcal{P} \label{eqn: 1c}\\[0.1cm]
& \sum_{g' \in \mathcal{P}_g}z_{g'} \leq C_g y_g  \quad \forall g \in \mathcal{G'} \label{eqn: 1d}\\[0.1cm]
& \sum_{g \in \mathcal{G} \cup \mathcal{P}} x_{wg} \leq 1 \quad \forall w \in \mathcal{W} \label{eqn: 1e}\\[0.1cm]
& \sum_{w \in \mathcal{W}} x_{wg}c_w + u_{g} =  C_g \quad \forall g \in \mathcal{G} \cup \mathcal{P} \label{eqn: 1f}\\[0.1cm]
& \sum_{w \in \mathcal{W}} x_{wg}m_w + S_g v_{g} = M_g \quad \forall g \in \mathcal{G} \cup \mathcal{P} \label{eqn: 1g}\\[0.1cm]
& y_{g'} + y_{g} \leq 1 \qquad \forall (g',g) \in \mathcal{B} \label{eqn: 1h} \\
& u_g - v_g \leq U_g \qquad \forall g \in \mathcal{G} \cup \mathcal{P}  \label{eqn: 1i} \\
& \delta_g \leq u_g \leq C_g \delta_g \qquad \forall g \in \mathcal{G} \cup \mathcal{P}  \label{eqn: 1k}\\
& v_g - M_g \delta_g \leq V_g \qquad \forall g \in \mathcal{G} \cup \mathcal{P} \label{eqn: 1j}  
\end{align}
\end{subequations}

The objective function (\ref{eqn: 1a}) is a profit maximization function that maximizes the gain from the workloads placed (term 1) while minimizing the costs related to GPUs that are used (term 2), the GPUs that need to be repartitioned (term 3), the workloads that are migrated (term 4) and lastly the wastage from unusable compute and memory slices (term 5). The reason behind the structure of some of these terms (e.g., term 4 and 5) will be explained below as we describe the remaining constraints and corresponding parameters and variables.  
By de-prioritizing migrations over saved GPUs through the choice of penalty, we ensure that workload migrations occur only if GPUs can be saved. Similarly by tuning other model weights,  we can prioritize one action over another, for example, placement over saving a GPU, and the latter over other costs.

Constraint (\ref{eqn: 1b}) ensures that if at least one workload is assigned to an unpartitioned GPU, then the corresponding $y_g = 1$ to indicate that the GPU is being used. In the scenario where a GPU has been partially partitioned (that is, it is in $\mathcal{G'}$), the feasible partition sizes, determined as per the aforementioned alignment constraints, will need to be respected while placing workloads on it. For this, we introduce a new auxiliary variable $z_g$ for each partition $g \in \mathcal{P}$  that takes the value of $1$ when a model $w$ is assigned to it, and $0$ otherwise. This aspect is handled jointly by constraints (\ref{eqn: 1c}) and (\ref{eqn: 1d}). 
Constraint (\ref{eqn: 1c}) ensures that the number of workloads placed in a feasible partition can be as many as the available compute, $C_{g'}$, on that partition. This means that in Figure~\ref{fig:DRA1} the 2g.20gb partition can either accommodate a single 2g.20gb workload or a single 1g.20gb workload or two 1g.10gb workloads.
Constraint (\ref{eqn: 1d}) ensures that the number of partitions on a GPU does not exceed the number of compute instances available on the GPU. For example, on a A100 containing 7 compute instances, there can be a maximum of 7 partitions each with 1g compute instances. Conversely, constraint (\ref{eqn: 1d}) ensures that if a GPU contains at least one partition with at least one workload placed in it, then the $y_g$ corresponding to such a partitioned GPU is set to 1. 
Constraint (\ref{eqn: 1e}) ensures that a workload may be placed on at most one unpartitioned GPU or on a feasible partition in an unpartitioned slice of a GPU. 
Constraint (\ref{eqn: 1f}) enforces the bin packing constraint on the compute instances dimension for each GPU $g$, that is, the total number of instances needed by the workloads assigned to a GPU should not exceed $C_g$,  with $u_g$ corresponding to the non-negative compute slice slack. 
Similarly, constraint (\ref{eqn: 1g}) enforces the bin packing constraint on the memory dimension for each GPU $g$, that is, the total memory required by the models assigned to a GPU should not exceed $M_g$, with $v_g$ corresponding to the non-negative memory slice slack.
Constraint (\ref{eqn: 1h}) ensures that if an imaginary GPU is used, as part of reconfiguration, then the original GPU is not active. This enforces that the maximum number of truly available GPUs for placement is always $|\mathcal{G'} \cup \mathcal{G} \setminus \mathcal{G}^{i}|$. If an imaginary GPU $g$ is part of the feasible solution, it means that the corresponding original GPU $g' \in \mathcal{G'}$ where $(g',g) \in \mathcal{B}$ has been reconfigured and the new loads on it are as per $g$. The migration term (4th term in the objective) can now be elaborated. A workload $w$ that was originally placed on $g'$ is viewed as migrated if it is neither on $g'$ nor on its imaginary counterpart $g$ (i.e., both $x_{wg'}$ and $x_{wg}$ are both zero). 

The last set of constraints aim to identify the wasted compute $U_g$ and memory $V_g$ slices to penalize a placement of that nature (see~Fig.~\ref{fig:gpu-waste}). 
Constraint (\ref{eqn: 1i}) together with the minimization objective over $U_g$ ensures that $U_g$ is precisely the  excess compute slices over memory slices, as $u_g$ and $v_g$ represent the slack in the respective resource constraints (in terms of slices). On the contrary, wastage in memory slices happen only when compute is full. To model the latter this, we introduce a binary variable, $\delta_g$ in constraint (\ref{eqn: 1k}) which ensures that $\delta_g$ is 1 if the compute is full and 0 otherwise. Constraint (\ref{eqn: 1j}) together with objective ensures the wasted memory slices $V_g$ are computed exactly.

\subsection{Rule-based / Heuristic Placement Model}
\label{s:ruleBasedModel}
Placement optimization for MIG workloads has three main use cases as explained in Section~\ref{s:usecases}. The MIP model described in the previous section solves all use cases jointly and finds the optimal placement for the defined objectives under the listed constraints. However, as bin-packing is an NP-hard problem, there maybe a significant computational overhead for large-sized problems. Moreover, its optimal solution may lead to sequential migration, which can further increase the execution time and possibly delay achieving the optimal state. 
To overcome these challenges, we propose to solve each use case separately by considering only the use case specific goals and constraints. Furthermore, we restrict our proposed solution to minimize sequential migration. This design by functional component for each use-case can thus be leveraged at different time horizons or cycles.

\textbf{Initial deployment} recommendations aim to place new workloads to  minimize the compute and memory wastage while maximizing utilization. 
\begin{itemize}
\item[Step 1:] Sort the new workloads in desending order of their size. We use profile id as a proxy for size since sorting by profile id also provides the sorted list for the size as shown in Table~\ref{tab:profiles}. 
\item[Step 2:] Sort the GPUs according to the sum of their joint memory and compute slice utilization. Mathematically, we define this joint slice utilization as $(s_m+s_c)/(S_m+S_c)$, where $s_m,s_c$ are the used memory and compute slices and $S_m, S_c$ are the total slices on the GPU. 
\item[Step 3:] Assign each workload to the GPU partition which provides maximum utilization after assignment. To decide the placement index, use the preference order in Table~\ref{tab:profiles}. If there is no available partition for the workload, allocate a new GPU to place the workload. 
\end{itemize}

The sorting of workloads also guarantees minimizing the memory wastage, since we prioritize profiles that use an extra memory slot (e.g. 3g.40gb, 1g.20gb in GPU slice 6, see Fig.~\ref{fig:gpu-waste}).

\textbf{Compaction} recommendations target vacating an underutilized GPU by migrating workloads to other GPUs with existing workloads. It shares the same goals and constraint with initial deployment component, i.e.,  to minimize the wastage and maximize utilization.
\begin{itemize}
\item[Step 1:] Sort the GPUs per the sum of their memory and compute slice utilization. 
\item[Step 2:] For each GPU starting from the least utilized, retrieve the existing workloads.
\item[Step 3:] Check the number of available slices on other allocated GPUs. If the available slices are enough to place the retrieved workloads, assign them to other GPUs by the process used in initial deployment which targets maximizing utilization.  
\end{itemize}

In some conditions, although the available GPUs have enough number of available slices for the workloads, they may not have availability for the right partitions for the workload. Sequential migration can solve this issue but in this heuristic approach we avoid sequential migration. Instead, we add a free GPU to the list of allocated GPUs and rerun Step 3. If Step 3 allows saving more than one GPU, it leads to a feasible compaction solution.

\begin{figure}[h]
  \centering
  \includegraphics[width=\linewidth]{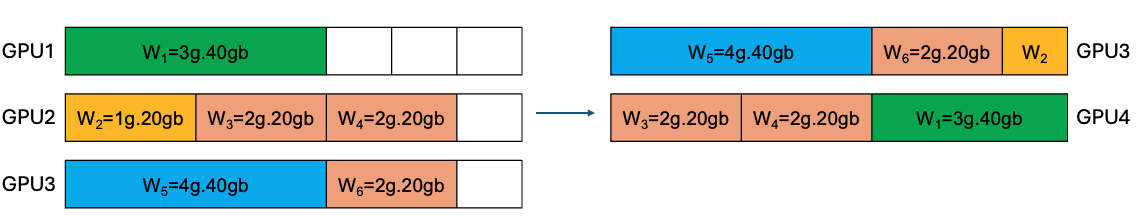}
  \caption{Compaction by using a free GPU.}
  \label{fig:compaction_with_free_gpu}
  \Description{}
\end{figure}

Figure~\ref{fig:compaction_with_free_gpu} demonstrates such a case that moving all workloads in any of the GPUs is not possible without sequential migration. To avoid sequential migration, we propose adding a new GPU and migrating all workloads in GPU1 and GPU2 to GPU3 and newly added GPU4. Moving workloads considering maximizing utilization allowed saving compute slice wastage in initial state due to $W_1$ and $W_2$. Eventually, it vacates 2 GPUs as opposed to allocating a new GPU and allows saving a GPU. 

\textbf{Reconfiguration / Redeployment} refers to achieving an optimal placement of all existing workloads when there are enough free GPUs to serve them. We consider this component is useful as part of maintenance cycles and assigned node changes such as node replacement or provider change. Reconfiguration targets using minimum number of GPUs for the given set of workloads by minimizing compute or memory wastage, maximizing utilization, and maximizing the availability of used GPUs for future workloads.
\begin{itemize}
\item[Step 1:] Calculate the minimum number of GPUs needed for placing all workloads using Equation~\ref{eq:min_gpu} where $\mathcal{W}$ is the set of all workloads, $c_w, m_w$ are the compute and memory slices needed for workload for $w \in \mathcal{W}$ and lastly $C$ and $M$ are the maximum number of compute slices and total memory size in a GPU respectively.
\begin{align}
\label{eq:min_gpu}
min\_GPUs = \max \Big\{\sum_{w \in \mathcal{W}} c_w/C, \sum_{w \in \mathcal{W}} m_w/M \Big \}
\end{align}
\item[Step 2:] Sort the GPUs by the sum of their compute and memory slice utilization and take the minimum number of GPUs from the least utilized GPUs.
\item[Step 3:] Assign workloads with additional memory profiles (first profile 9 and then profile 15) first until either all workloads are assigned or there is one for each GPU. This ensures minimizing memory wastage on GPUs.
\item[Step 4:] Sort all remaining workloads by their profile ids which gives the sorted order of workloads by their GPU slices need.
\item[Step 5:] Place workloads in GPUs found in Step 2 by using first-fit bin packing algorithm. Before each placement decision, check the feasibility of the assignment by using allowed indexes listed in Table~\ref{tab:profiles}. 
\end{itemize}

If there are enough free GPUs to place all workloads, Step 2 ensures using only free GPUs. If not, it uses least utilized GPUs to minimize sequential migration need. At Step 3, we first place all workloads with Profile 9 since it requires more GPU slices than Profile 15. Sorting workloads by their size at Step 4 converts GPU placement optimization to a one-dimensional bin-packing problem since it considers GPU slices needed for each workload. At Step 5, we use first-fit bin backing algorithm for decreasing workload sizes. At each workload-GPU assignment, we check the feasibility by checking the current placement state of the assigned GPU and the availability of the allowed indexes and consecutive slices of GPU for the workload size using Table~\ref{tab:profiles}. To achieve the maximum availability and decide the right index of a workload, we employ the preference order of GPU slices listed in the table. 

We want to point that in Table~\ref{tab:profiles} the allowed indices are not given in increasing order. We empirically identified the preference order with our experiments using nvidia-smi package. For instance, for an empty GPU if a workload with profile 15 comes, since all slices are available it will be located at index 6. However, if slice 6 was not available, it would be placed at slice 4 and utilize both slice 4 and 5 due to 2 memory slices requirement. It could also be placed at slice 0 or 2 and still use 2 slices. However, it would decrease the availability of the GPU for other workloads with profile 5 or 9 as also explained in Section~\ref{s:maximizeAvailability}. Finally, if a feasible bin-packing solution is not found for the given set of workloads and GPUs, minimum GPU count is increased by one and Step 2-5 are re-executed for the new set of GPUs.

\section{Evaluation}
  
We now describe the experimental setup, metrics and methodology and present the results.  
All approaches are implemented in Python and we use IBM CPLEX 22.1.1~\cite{IBM_CPLEX} to solve the MIP.

\subsection{Experimental Setup}
\label{s:setup}
We use a simulation framework that mimics a cluster with one or more GPU nodes where each node has eight GPUs. We consider a homogeneous cluster for clarity but the proposed approaches can address placement in clusters with heterogeneous GPU types. For our experiments, we considered two different cluster sizes: a single node (8 GPUs) and ten nodes (80 GPUs). For each cluster size, we generated 100 test cases, that is 200 test cases totally, on which we evaluated all our approaches and use cases.  

The MIP models were solved to optimality on test-cases with 8 GPUs. For test-cases with 80 GPUs, the optimizer quickly converges to within 0.2\% of optimality in 1-2 seconds, but takes longer to converge to the global optimal (as is the case always, generating a certificate of optimality takes time compared to solution convergence). Hence we run experiments for the 80 GPUs by setting CPLEX time limit to 30 seconds.

\textbf{Test cases:} For each test case, we allocated about 60\% of GPUs and left the remaining GPUs free. For each GPU, utilization (up to 100\%) is randomly assigned. 
Next, workloads and their profiles to achieve the assigned GPU utilization are both randomly generated from the profiles listed in Table~\ref{tab:profiles}. This way, we provide a random initial state for a GPU cluster with some GPUs free and others allocated with a diverse set of workloads. In addition, for the initial deployment case, we also generated random workloads. The total size of the generated workloads corresponds to the 60\% of the total capacity of the cluster to ensure that the cluster would be highly allocated after initial deployment. 

\textbf{Approaches:} We benchmark the performance of our rule-based and MIP approaches with two other commonly used workload scheduling algorithms: first-fit and resource-based dynamic load-balancing algorithm.
The first-fit algorithm initially sorts GPUs and workloads by their respective IDs, and then assigns each workload to the available partitions on the ordered set of GPUs by starting at index 0 for each GPU.
The resource-based dynamic load-balancing algorithm sorts the GPUs by their total compute and memory slice utilization in the ascending order.
Then, it processes each workload in the order of which they are received and assigns them to the available partitions on the ordered set of GPUs, starting from index 0.
We check the feasibility of the placement decisions at each step, similar to our rule-based approach, to ensure that only feasible placements are allowed.

\textbf{Metrics:}
The final placement solution for each use-case from each approach is evaluated for number of GPUs used, memory wastage, compute wastage, availability, migration size, pending approach size, memory utilization, compute utilization, and whether or not the final placement requires sequential migration. Table~\ref{tab:metrics} defines each metric used for our evaluations.

\begin{table}
    \small
    \centering
    \rowcolors{1}{}{gray!10}
    \begin{tabular}{l p{0.3\textwidth}}
    \toprule
    {\bf Metric} & {\bf Definition}\\
    \midrule
        \# GPUs & Total number of GPUs with at least one workload in the final placement \\
        \midrule
        Memory Wastage & Total number of memory slices wasted in the final placement (here due to placing a workload at index 6 with profile 19 
        )\\
          \midrule
        Compute Wastage & Total unusable compute slices in the final placement (here due to placing profile 9 workload at index 0, or profile 15 at any index other than 6)\\
          \midrule
        Availability  & Number of available GPU slices remaining on the cluster after all workloads are placed. If all workloads are not placed, availability is calculated by subtracting the total pending workload size from the number of available GPU slices\\
          \midrule
        Migration Size & Total memory size of existing workloads moved to new GPUs in the final placement\\
          \midrule
        Pending Model Size & Total number of memory slices needed to place workloads that remain in a pending state in the final placement 
        \\
          \midrule
        Sequential Migration & Number of existing workloads that need sequential migration to be placed as per the final placement solution. Sequential migration is needed when a workload is moved to another GPU in the final state, and a suitable partition is not available at that index in the initial state.\\
          \midrule
        Memory Utilization & Percentage of total memory slices used for final placement (ratio of used memory slices to the total available memory slices in the used GPUs)\\
          \midrule
        Compute Utilization & Percentage of total compute slices used for final placement (ratio of the used compute slices to the total available compute slices in used GPUs.) \\
        \bottomrule
    \end{tabular}  
    \caption{Metrics observed and reported}
    \label{tab:metrics}
\end{table}

\subsection{Evaluation Results}
\label{s:evaluationResults}
All evaluations use the average results of 100 test cases for each cluster and use case and provided with a normalized value against the highest value for each metric over all tested approaches. Therefore, it shows the relative performance of each approach for the listed metrics. 

\begin{figure*}
\centering
\begin{subfigure}{1.0\textwidth}
  \centering
  \includegraphics[width=1.0\linewidth]{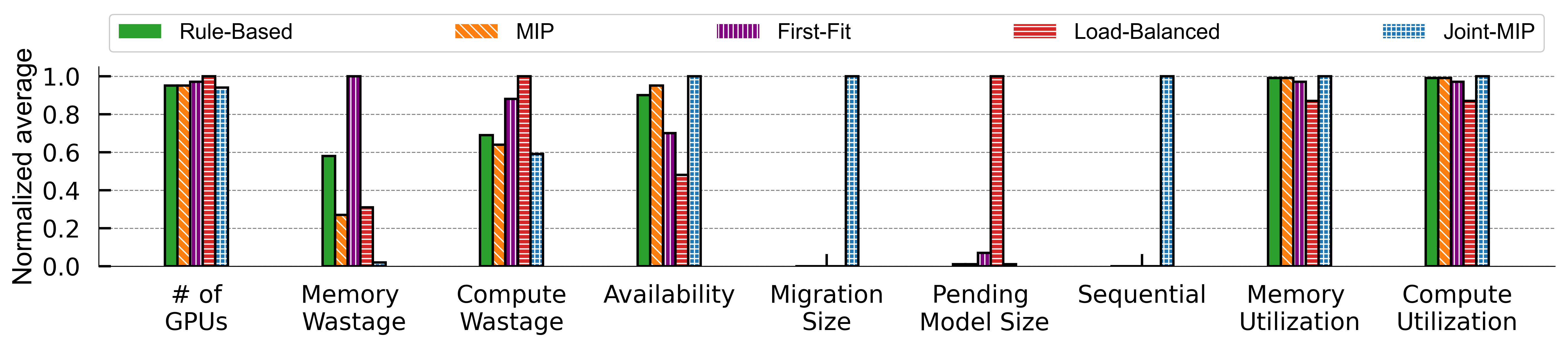}
  \caption{Average results for a cluster with 8 GPUs.}
  \label{fig:initial_8}
\end{subfigure}
\hfill
\begin{subfigure}{1.0\textwidth}
  \centering
  \includegraphics[width=1.0\linewidth]{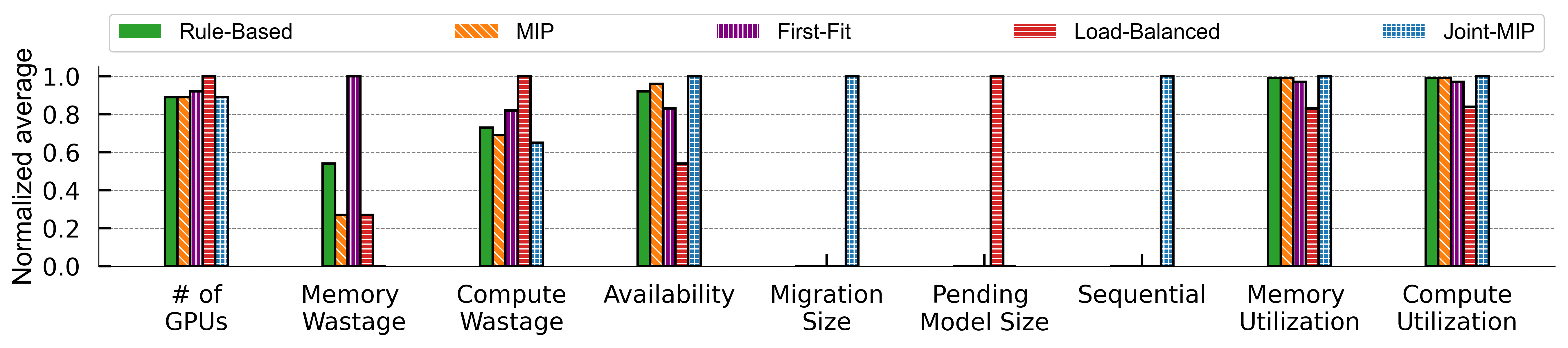}
  \caption{Average results for a cluster with 80 GPUs.}
  \label{fig:initial_80}
\end{subfigure}
\caption{Comparison of placement approaches for initial deployment use case for two cluster sizes. 
}
\label{fig:initial_deployment_evaluations}
\end{figure*}
\subsubsection{Initial Deployment Experiments}
\label{s:exp_initial_deployment}
All experiments were run using both existing and new workloads, and partitioned and new (unused) GPUs. For \emph{rule-based}, \emph{MIP}, \emph{first-fit} and \emph{load balanced}, we fix the placement of existing workloads as provided by their initial state, and allow the approach to place new workloads on all GPUs. For \emph{joint-MIP}, we relax this restriction and allow the MIP to find the optimal placement for both existing and new workloads. That is, \emph{joint-MIP} effectively handles initial deployment of new workloads, and compaction and reconfiguration in terms of optimally placing the existing workloads. As a result, \emph{joint-MIP} is expected to outperform the others on the number of GPUs used, wastage, availability and utilization metrics. This causes it to have higher migrations and sequential steps w.r.t. the other approaches.

Figure~\ref{fig:initial_deployment_evaluations} shows the evaluation results for initial deployment use case.  
For initial deployment of new workloads, the number of GPUs used in the final placement solution and the size of pending workloads are the critical metrics to track. Memory and compute utilization show how efficiently the new workloads were placed. For the number of GPUs used, we observed a 6\% improvement in \emph{joint-MIP} w.r.t. \emph{load-balanced} and a 5\% improvement in \emph{MIP} and \emph{rule-based}  w.r.t. \emph{load-balanced} for cluster with 8 GPUs while avoiding any pending workloads. The performance improvement of these models become more significant for cluster with 80 GPUs. \emph{Joint-MIP},\emph{MIP}, and \emph{rule-based} all achieves 11\% improvement compared to \emph{load-balanced}. Additionally, in spite of using higher number of GPUs, \emph{load-balanced} has pending workloads in every test-case for both cluster sizes, whereas \emph{MIP} and \emph{rule-based} have one pending workload in a single test-case with 8 GPUs. \emph{First-fit} produces pending workloads in 7 test-cases with 8 GPUs. The performance of \emph{joint-MIP} is in line with our expectations stated above. \emph{Rule-based} and \emph{MIP} achieve 99\% of \emph{joint-MIP} performance for memory and compute utilization for both cluster sizes, while they achieve at least 90\% of \emph{joint-MIP} performance for availability. In terms of wastage, we note that memory wastage is lower for \emph{load-balanced} compared to \emph{rule-based}. However, its performance on this metric is undercut by the fact that it has pending workloads for every test-case.

\subsubsection{Compaction Experiments}
\label{s:exp_compaction}

\begin{figure*}
\centering
\begin{subfigure}{1.0\textwidth}
  \centering
  \includegraphics[width=1.0\linewidth]{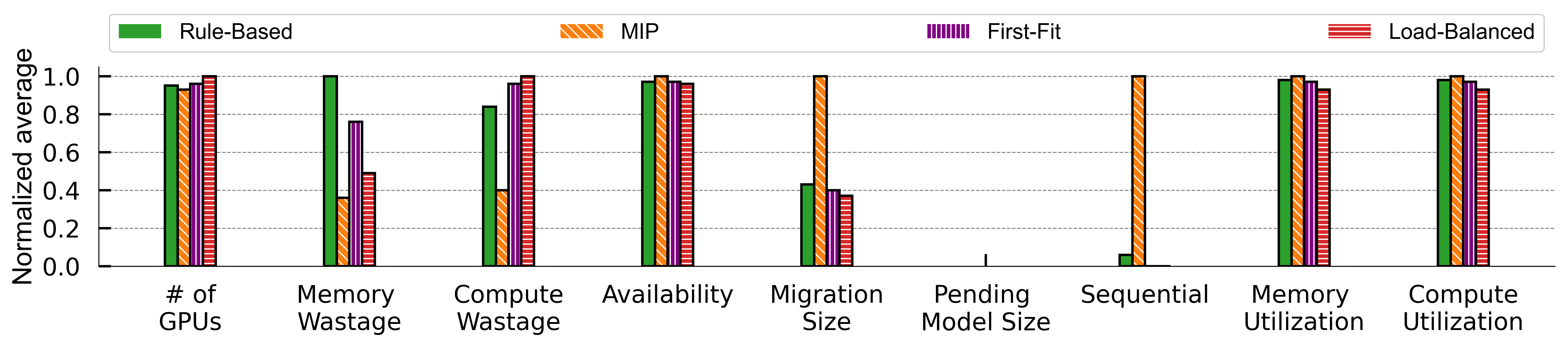}
  \caption{Average results for a cluster with 8 GPUs.}
  \label{fig:compaction_8}
\end{subfigure}
\hfill
\begin{subfigure}{1.0\textwidth}
  \centering
  \includegraphics[width=1.0\linewidth]{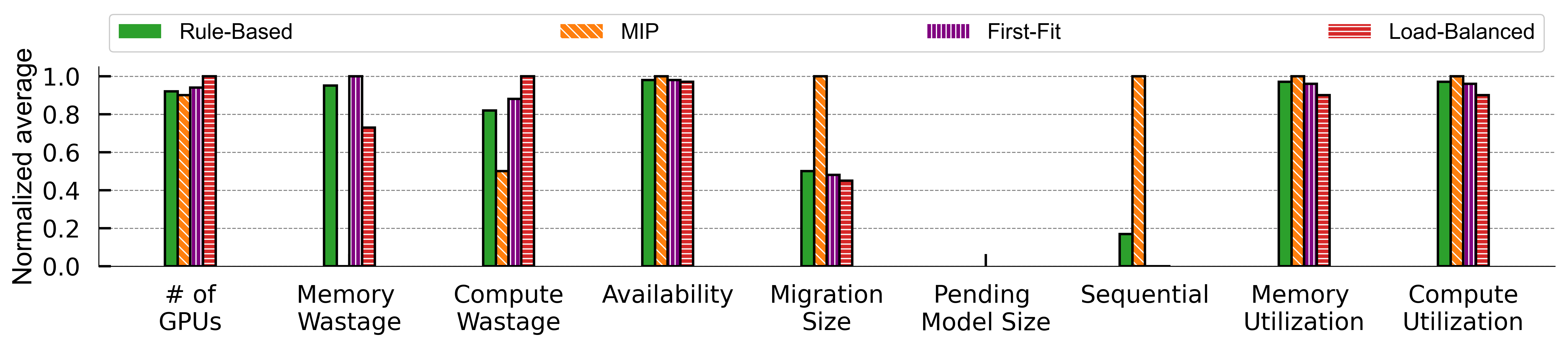}
  \caption{Average results for a cluster with 80 GPUs.}
  \label{fig:compaction_80}
\end{subfigure}
\caption{Comparison of placement approaches for compaction use case for two cluster sizes. 
}
\label{fig:compaction_evaluations}
\end{figure*}

For compaction, we consider only existing workloads. As a result, pending workloads are null for all methods. The key metrics here are number of GPUs used, wastage and migration size. Both of our methods outperform the others in terms of number of GPUs used, with \emph{MIP} having a 2\% improvement over \emph{rule-based}. Also, \emph{MIP} improves the number of GPUs by 7\% for 8 GPUs and 10\% for 80 GPUs compared to \emph{load-balanced} while \emph{Rule-based} achieves 5\% and 8\% improvements respectively. We observe that \emph{MIP} has a higher number of migrations that need sequential steps. \emph{MIP} outperforms in terms of memory and compute wastage as well. Importantly, we observed that \emph{load-balanced} has lower memory wastage than \emph{rule-based}. This is because slice at index 6 is used less frequently by \emph{load-balanced} in the test-cases.

\subsubsection{Reconfiguration Experiments}
\label{s:exp_reconfiguration}

\begin{figure*}
\centering
    \begin{subfigure}{1.0\textwidth}
      \centering
      \includegraphics[width=1.0\linewidth]{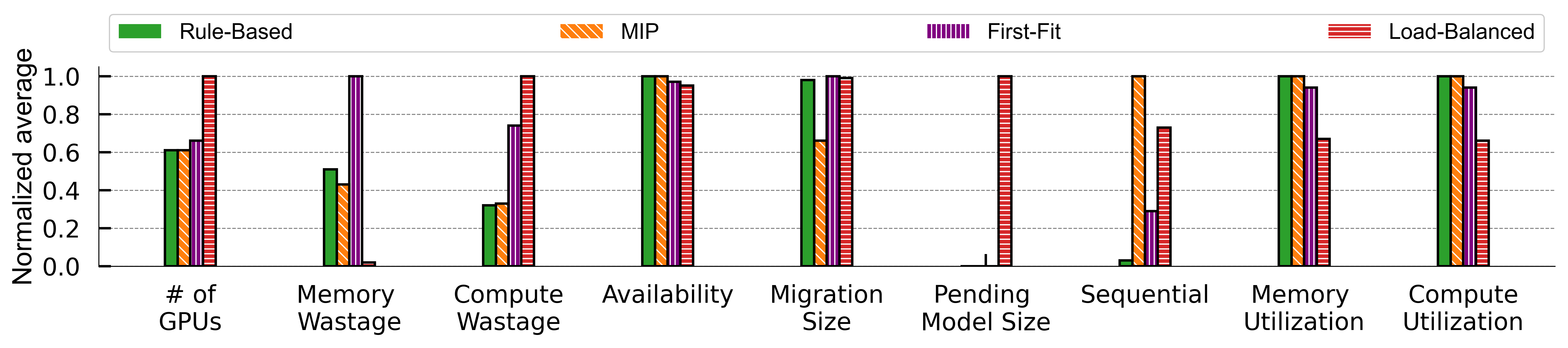}
      \caption{Average results for a cluster with 8 GPUs.}
      \label{fig:reconfiguration_8}
    \end{subfigure}
    \hfill
    \begin{subfigure}{1.0\textwidth}
      \centering
      \includegraphics[width=1.0\linewidth]{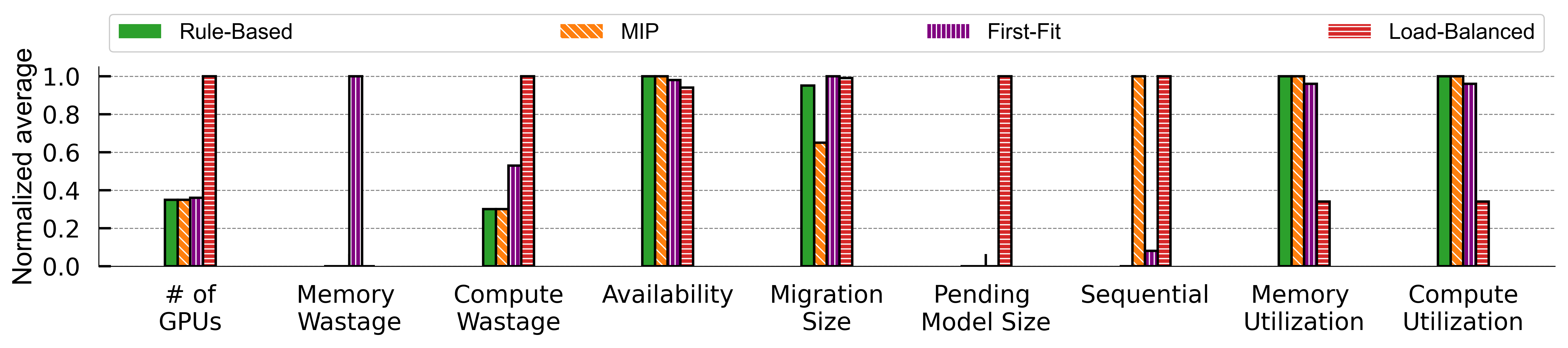}
      \caption{Average results for a cluster with 80 GPUs.}
      \label{fig:reconfiguration_80}
    \end{subfigure}
\caption{Comparison of placement approaches for reconfiguration use case for two cluster sizes. 
}
\label{fig:reconfiguration_evaluations}
\end{figure*}
In reconfiguration, we consider both used and new GPUs but only existing workloads. The intent is to place as many existing workloads on new GPUs and, in the process, determine the best set of partition indices for the new GPUs so that we minimize the number of GPUs used and wastage. If all existing workloads cannot be placed on new GPUs, then we allow placing workloads on used GPUs, and repartition them if necessary. Ultimately, the number of migrations to achieve the final placement needs to be minimized.  

Both \emph{MIP} and \emph{rule-based} achieve similar performance, with the \emph{MIP} doing slightly better on memory wastage and migration size. Once again, the \emph{MIP} placement solution needs higher number of sequential steps. \emph{MIP} and \emph{rule-based} achieve 39\% improvement in number of GPUs compared to \emph{load-balanced} for 8 GPUs and up to 65\% improvement for 80 GPUs. Additionally, our methods are able to place all workloads and achieving better compute and memory wastage. \emph{Load-balanced} assigns workloads one by one for the available GPUs starting from index 0, which leads to having some workloads with profile 5 to stay in pending state since they can only be placed at index 0.

\section{Related Work}
GPU resources are scarce considering the increasing demand in generative models. Similarly, the cost of using GPUs is considerably higher than other compute and memory resources.
Using a GPU for a single job commonly results in under-utilization of GPU resources. In this regard, existing works propose sharing GPU resources for multiple tasks. 

Espenshade et al.~\cite{10.1145/3642970.3655830} present measurements highlighting how to tune a large modern GPU to run workloads that may not need an entire GPU.
They improve the aggregate throughput and efficiency of the GPU by placing smaller workloads onto spatially partitioned GPUs using MIG.
MISO \cite{Li_2022} proposes predicting the best MIG partition for different workloads by leveraging the Multi-Process Service (MPS) capability to predict the best MIG partition.
Similarly, Adufu et al.~\cite{10572198} combine MPS (Multi-Process Service) and MIG capabilities from Nvidia to address the issues pertaining to mitigate resource under-utilization.
Finally, Choi et al.\cite{choi2021multimodelmachinelearninginference} propose a new abstraction for GPUs called \textit{gpu-lets}, which leverages the spatial partitioning feature\textemdash MPS capability on Nvidia GPUs. 
Then they leverage \textit{gpu-lets} to allocate GPU resources to jobs with the goal of increasing resource utilization.Despite the interest in GPU resource optimization, none of these works target optimizing MIG partition placement. 

\section{Conclusion and Future Work}

Our results show up to 2.85x improvement in the number of GPUs used and up to 70\% reduction in GPU wastage over baseline heuristics.  
Encouraged by these improvements, we plan to deploy our work on production systems and enable SREs to consume recommendations from our framework via a command-line tool.
The framework in Figure \ref{fig:framework} takes into account the existing configuration of GPUs in the IT environment.
We also plan to make reconfiguration recommendations to the SREs for their environments every two weeks\textemdash aligning with their maintenance windows.
This would then allow us to request explicit feedback and improve on the recommendations made.
While at this time we expect the SREs to manually act on these recommendations, in the near future we plan to leverage InstaSlice\cite{instalice-github} as an actuation mechanism.
InstaSlice facilitates the allocation of MIG slices on Nvidia GPUs. It acts on queued AI workloads to slice GPUs with the help of DRA and places them with no changes needed to the queued workloads or to the Kubernetes schedulers.

We intend to undertake the following extensions:
    (1) Draw-up sequential migration plans (when applicable)
    (2) Develop a method to estimate the real costs of migrations and re-partitioning
    (3) Create an approach for MIG allocation of fine-tuning workloads.

\bibliographystyle{ACM-Reference-Format}
\bibliography{main}
\end{document}